# Quantum criticality in an Ising chain: experimental evidence for emergent E8 symmetry

R. Coldea[1,*], D.A. Tennant[2], E.M. Wheeler[1,†], E. Wawrzynska[3], D. Prabhakaran[1], M. Telling[4], K. Habicht[2], P. Smeibidl[2], K. Kiefer[2]

[1]Clarendon Laboratory, University of Oxford Department of Physics, Parks Road, Oxford OX1 3PU, United Kingdom

[2†]Helmholtz-Zentrum Berlin für Materialien und Energie, Lise Meitner Campus, Glienicker Str. 100, D-14109 Berlin, Germany

[3]H.H. Wills Physics Laboratory, University of Bristol, Tyndall Avenue, Bristol, BS8 1TL, United Kingdom

[4]ISIS, Rutherford Appleton Laboratory, Chilton, Didcot, OX11 0QX, United Kingdom

[†]present address

[*]E-mail: r.coldea@physics.ox.ac.uk

**Quantum phase transitions take place between distinct phases of matter at zero temperature. Near the transition point, exotic quantum symmetries can emerge that govern the excitation spectrum of the system. A symmetry described by the E8 Lie group with a spectrum of 8 particles was long predicted to appear near the critical point of an Ising chain. We realize this system experimentally by tuning the quasi-one-dimensional Ising ferromagnet $CoNb_2O_6$ through its critical point using strong transverse magnetic fields. The spin excitations are observed to change character from pairs of kinks in the ordered phase to spin-flips in the paramagnetic phase. Just below the critical field, the spin dynamics shows a fine structure with two sharp modes at low energies, in a ratio that approaches the golden mean as predicted for the first two meson particles of the E8 spectrum. Our results demonstrate the power of symmetry to describe complex quantum behaviours.**

Symmetry is present in many physical systems and helps uncover some of their fundamental properties. Continuous symmetries lead to conservation laws, for example invariance of physical laws under spatial rotations ensures the conservation of angular momentum. More exotic continuous symmetries have been predicted to emerge in the proximity of certain quantum phase transitions (QPTs)(*1,2*). Recent experiments on quantum magnets (*3,4,5*) suggest that quantum critical resonances may expose underlying symmetries most clearly. Remarkably, the simplest of systems, the Ising chain, promises a very complex symmetry, described mathematically by the $E_8$ Lie group (*2,6-9*). Lie groups describe continuous symmetries and are important in many areas of physics. They range in complexity from the U(1) group, which appears in the low-energy description of superfluidity, superconductivity and Bose-Einstein condensation (*10,11*) to $E_8$, the highest-order symmetry group discovered in mathematics (*12*), which has not yet been experimentally realized in physics.

The one-dimensional (1D) Ising chain in transverse field (*10,11,13*) is perhaps the most studied theoretical paradigm for a quantum phase transition. It is described by the Hamiltonian

$$H = \Sigma_i - J S^z_i S^z_{i+1} - h S_i^x, \qquad (1)$$

where a ferromagnetic exchange $J > 0$ between nearest-neighbor spin-1/2 magnetic moments $S_i$ arranged on a one-dimensional chain competes with an applied external transverse magnetic field $h$. The Ising exchange $J$ favors spontaneous magnetic order along the z-axis ($|\uparrow\uparrow\uparrow \cdots \uparrow\rangle$ or $|\downarrow\downarrow\downarrow \cdots \downarrow\rangle$), in contrast the transverse field $h$ forces the spins to point along the perpendicular $+x$ direction ($|\rightarrow\rightarrow\rightarrow \cdots \rightarrow\rangle$). This competition leads to two distinct phases, magnetic order and quantum paramagnet, separated by a continuous transition at the critical field $h_C=J/2$ (Fig. 1A). Qualitatively, the magnetic field stimulates quantum tunneling processes between ↑ and ↓ spin states and these zero-point quantum fluctuations "melt" the magnetic order at $h_C$ (*10*).

To explore the physics of Ising quantum criticality in real materials several key ingredients are required: very good one-dimensionality of the magnetism to avoid mean-field effects of higher dimensions, a strong easy-axis (Ising) character and a sufficiently low exchange energy $J$ of a few meV that can be matched by experimentally-attainable magnetic fields (10 T~1 meV) to access the quantum critical point. An excellent model system to test this physics is the insulating quasi-one-dimensional Ising ferromagnet $CoNb_2O_6$ (*14-16*), where magnetic $Co^{2+}$ ions are arranged into near-isolated zig-zag chains along the *c*-axis with strong easy-axis anisotropy due to crystal field effects from the distorted $CoO_6$ local environment (Fig 1B). Large single crystals can be grown (*17*) and this is essential to be able to measure the crucial spin dynamics with neutron scattering.

$CoNb_2O_6$ orders magnetically at low temperatures below $T_{N1}$=2.95 K, stabilized by weak interchain couplings. The chains order ferromagnetically along their length with magnetic moments pointing along the local Ising direction, contained in the crystal (*ac*) plane (*18*). To tune to the critical point, we apply an external magnetic field along the *b*-axis, transverse to the local Ising axis. Fig. 1C shows that the external field suppresses the long-range 3D magnetic order favoured by the Ising exchange in a continuous phase transition at a critical field $B_C$= 5.5 T.

Expected excitations for the model in Eq. 1 consist of i) pairs of kinks (with the cartoon representation $|\uparrow\uparrow\downarrow\downarrow_z...\rangle$) below $B_C$ and ii) spin flips $|\rightarrow\rightarrow\leftarrow\rightarrow_x ...\rangle$ above $B_C$. The kinks interpolate between the two degenerate ground states with spontaneous magnetization along the +$z$ or –$z$ axis, respectively. Neutrons scatter by creating a pair of kinks (Fig. 2A). The results in Fig. 2B and C show that in the ordered phase below $B_C$ the spectrum is a bow-tie shaped continuum with strongly-dispersive boundaries and large bandwidth at the zone centre (L=0), which we attribute to the expected two-kink states. This continuum increases in bandwidth and lowers its gap with increasing field, as the applied transverse field provides matrix elements for the kinks to hop, directly tuning their kinetic energy. Above $B_C$ a very different spectrum emerges (Fig. 2E), dominated by a single sharp mode. This is precisely the signature of a quantum paramagnetic phase. In this phase the spontaneous ferromagnetic correlations are absent, and there are no longer two equivalent ground states that could support kinks. Instead, excitations are single spin reversals opposite to the applied field that cost Zeeman energy in increasing field. The

fundamental change in the nature of quasiparticles observed here (compare Fig 2C and E) does not occur in higher-dimensional realizations of the quantum Ising model. The kinks are a crucial aspect of the physics in one dimension and their spectrum of confinement bound states near the transition field will be directly related to the low-energy symmetry of the critical point.

The very strong dimensionality effects in 3D systems stabilize sharp spin-flip quasiparticles in both the ordered and paramagnetic phases, as indeed observed experimentally in the 3D dipolar-coupled ferromagnet LiHoF$_4$ (*19*). In contrast, weak additional perturbations in the 1D Ising model, in particular a small longitudinal field $-h_z \Sigma_i S_i^z$ should lead to a rich structure of bound states (*6,7,9*). Such a longitudinal field in fact arises naturally in the case of a quasi-1D magnet, as in the 3D magnetically-ordered phase at low temperature the weak couplings between the magnetic chains can be replaced in a first approximation by a local, effective longitudinal mean field (*21*), which scales with the magnitude of the ordered moment $\langle S^z \rangle$ [$h_z = \Sigma_\delta J_\delta \langle S^z \rangle$ where the sum extends over all interchain bonds with exchange energy $J_\delta$]. If the 1D Ising chain is precisely at its critical point ($h = h_C$) then the bound states stabilized by the additional longitudinal field $h_z$ morph into the "quantum resonances" that are a characteristic fingerprint of the emergent symmetries near the quantum critical point. Nearly two decades ago Zamolodchikov (*2*) proposed precisely eight "meson" bound states (the kinks playing the role of quarks), with energies in specific ratios given by a representation of the E$_8$ exceptional Lie group (*2*). Before discussing the results near the QPT, we develop below a more sophisticated model of the magnetism in CoNb$_2$O$_6$ including confinement effects at zero field, where conventional perturbation theories are found to hold.

The zero field data in Fig. 3A shows a gapped continuum scattering at the ferromagnetic zone centre (L=0) due to kink pairs, which are allowed to propagate even in the absence of an external field. This is caused by sub-leading terms in the spin Hamiltonian. Upon cooling to the lowest temperature of 40 mK, deep in the magnetically-ordered phase, the continuum splits into a sequence of a sharp modes (Fig. 3B). At least 5 modes can be clearly observed (Fig. 3E), and they exist over a wide range of wavevectors and have a quadratic dispersion (open symbols in Fig. 3D). This data demonstrates the physics of kink confinement under a linear attractive interaction (*6-9*). In the ordered phase, kink propagation upsets the bonds with the neighbouring chains (Fig. 3G) and therefore requires an energy cost $V(x)$ that grows linearly with the kink separation $x$, $V(x)= \lambda|x|$ where the "string tension" $\lambda$ is proportional to the ordered moment magnitude $\langle S^z \rangle$ and the interchain couplings strength [$\lambda=2h_z\langle S^z \rangle/\tilde{c}$, where $h_z=\Sigma_\delta J_\delta \langle S^z \rangle$ is the longitudinal mean field of the interchain couplings and $\tilde{c}=c/2$ is the lattice spacing along the chain].

The essential physics of confinement is apparent in the limit of small $\lambda$ for two kinks near the band minimum, where the one-kink dispersion is quadratic $\varepsilon(k)= m_o + \hbar^2 k^2/(2\mu)$. In this case, the Schrödinger's equation for the relative motion of two kinks in their centre-or-mass frame is (*6-9,22*)

$$-\hbar^2/\mu \, d^2\varphi/dx^2 + \lambda|x| \varphi = (m - 2m_o) \varphi, \qquad (2)$$

which has only bound state solutions with energies (also called masses)

$$m_j = 2m_o + z_j \, \lambda^{2/3} \, (\hbar^2/\mu)^{1/3}, \qquad j=1,2,3,\ldots \qquad (3)$$

The bound states are predicted to occur above the threshold $2m_o$ for creating two free kinks in a specific sequence given by the prefactors $z_n$, the negative zeros of the Airy function Ai($-z_n$)=0, $z_j$= 2.33, 4.08, etc. (*18*). The very non-trivial sequencing of the spacing between levels at the zone centre agrees well with the measured energies of all 5 observed bound states (Fig. 3H), indicating that the weak confinement limit captures the essential physics.

A full modelling of the data throughout the Brillouin zone can be obtained (*18*) by considering an extension of Eq. 2 to finite wavevectors and adding a short-range interaction between kinks, responsible for stabilizing the observed bound state near the zone boundary L=-1. Interestingly, this is a kinetic bound state, i.e. it is stabilized by the fact that two kinks gain extra kinetic energy if they hop together due to their short range interaction, as opposed to the Zeeman ladder of confinement bound states (near L=0), stabilized by the potential energy $V(x)$. The good agreement with the dispersion relations of all the bound states observed (Fig. 3D) as well as the overall intensity distribution (compare Fig. 3B and F) shows that an effective model of kinks with a confinement interaction can quantitatively describe the spin dynamics.

Having established the behaviour at zero field, we now consider the influence of the QPT at high field. Fig. 4C shows the excitation gap decrease upon approaching the critical field, as quantum tunnelling lowers the energy of the kink quasiparticles, then increase again above $B_C$ in the paramagnetic phase due to the increase in Zeeman energy cost for spin-flip quasiparticles. In a quasi-one-dimensional system such as $CoNb_2O_6$ with finite interchain couplings, a complete gap softening is only expected (*23*) at the location of the 3D magnetic long-range order Bragg peaks, which occur at a finite interchain wavevector $q_\perp$ that minimises the Fourier transform of the antiferromagnetic interchain couplings; the measurements shown in Fig. 4C were in a scattering plane where no magnetic Bragg peaks occur, so an incomplete gap softening would be expected here, as indeed observed.

For the critical Ising chain a gapless spectrum of critical kinks is predicted (Fig. 4F). Adding a finite longitudinal field $h_z$ generates a gap and stabilizes bound states (Fig. 4G). In the scaling limit sufficiently close to the quantum critical point, i.e. $h_z \ll J$, $h=h_C$, the spectrum is predicted to have 8 particles with energies in specific ratios (given by a representation of the E8 Lie group) with the first mass at $m_1/J = C(h_z/J)^{8/15}$, $C \approx 1.59$ (*2*). The predicted spectrum for such an off-critical Ising chain to be observed by neutron scattering is illustrated in Fig. 4E for the dominant dynamical correlations $S^{zz}(k=0,\omega)$ for which quantitative calculations are available (*7*): two prominent sharp peaks due to the first two particles $m_1$ and $m_2$ are expected at low energies below the onset of the continuum of two $m_1$ particles (*24*).

Figs. 4A and B show the neutron data taken just below the critical field. This is indeed consistent with this highly non-trivial prediction of two prominent peaks at low energies, which we identify with the first two particles $m_1$ and $m_2$ of the off-critical Ising model. Fig. 4D shows how the ratio of the energies of those peaks varies with increasing field and approaches closely (near 5 T just below the 3D critical field of 5.5 T) the golden ratio $m_2/m_1 = (1+\sqrt{5})/2 = 1.618$ predicted for the $E_8$ masses. We identify the field where the closest agreement with the $E_8$ mass ratio is observed as the field $B_C^{1D}$ where the 1D chains would have been critical in the absence of interchain couplings (*25*). Indeed, it is in this regime (*21*) that the special quantum critical symmetry theory would be expected to apply.

Our results emphasize that exploration of continuous quantum phase transitions can open up avenues to realize experimentally otherwise inaccessible (*1,26*) correlated quantum states of matter with complex symmetries and dynamics.

**References and Notes**

[1] For a review see F.H.L. Essler and R.M. Konic, arXiv:cond-mat/0412421 (2004).

[2] A.B. Zamolodchikov, *Int. J. Mod. Phys.* **A4**, 4235 (1989).

[3] B. Lake, D. A. Tennant, and S. E. Nagler, *Phys. Rev. Lett.* **85**, 832 (2000).

[4] M. Kenzelmann, Y. Chen, C. Broholm, D. H. Reich, and Y. Qiu, *Phys. Rev. Lett.* **93**, 017204 (2004).

[5] Ch. Rüegg *et al.*, *Phys. Rev. Lett.* **100**, 205701 (2008).

[6] B.M. McCoy and T. T. Wu, *Phys. Rev. D* **18**, 1259 (1978).

[7] G. Delfino, G. Mussardo, *Nucl. Phys.* **B455**, 724 (1995)

[8] G. Delfino, G. Mussardo, P. Simonetti, *Nucl. Phys.* **B473**, 469 (1996).

[9] P. Fonseca, A. Zamolodchikov, arXiv:hep-th/0612304 (2006).

[10] S. Sachdev, Quantum Phase Transitions, (Cambridge Univ. Press, Cambridge, 1999).

[11] A.O. Gogolin, A.A. Nersesyan and A.M. Tsvelik, Bosonization and Strongly Correlated Systems (Cambridge Univ. Press, Cambridge, 1998).

[12] D. Vogan, *Notices of the AMS* **54,** 1022 (2007).

[13] P. Pfeuty, *Ann. Phys. (N.Y.)* **57**, 79 (1970).

[14] C. Heid, H. Weitzel, P. Burlet, *et al.*, *J. Magn. Magn. Mater.* **151**, 123 (1995).

[15] S. Kobayashi *et al*, *Phys. Rev. B* **60**, 3331 (1999).

[16] I. Maartense, I. Yaeger, B.M. Wanklyn, *Solid State Comm.* **21**, 93 (1977).

[17] D. Prabhakaran, F.R. Wondre, and A.T. Boothroyd, *J. Crystal Growth* **250**, 72 (2003).

[18] Materials and methods are available as supporting material on Science online.

[19] H.M. Rønnow *et al*, *Science* **308**, 389 (2005).

[20] D. Bitko, T. F. Rosenbaum, and G. Aeppli, *Phys. Rev. Lett.* **77**, 940 (1996).

[21] S.T. Carr and A.M. Tsvelik, *Phys. Rev. Lett.* **90**, 177206 (2003).

[22] S.B. Rutkevich, *J. Stat. Phys.* **131**, 917 (2008).

[23] S. Lee, R.K. Kaul and L. Balents (unpublished).

[24] The higher energy particles $m_3$-$m_8$ are expected to produce much smaller features in the total scattering lineshape as they carry a much reduced weight and are overlapping or are very close to the lower boundary onset of the continuum scattering, see Fig 4E.

[25] The small offset between the estimated 1D and 3D critical fields is attributed to the interchain couplings, which strengthen the magnetic order. We note that a more precise quantitative comparison with the E8 model would require extension of the theory to include how the mass ratio $m_2/m_1$ depends on interchain wavevector $q_\perp$, as the data in Fig. 4D was collected slightly away from the 3D Bragg peak positions; the already good agreement with the long-wavelength prediction expected to be valid near the 3D Bragg wavevector may suggest that the mass ratio dispersion is probably a small effect at the measured wavevectors.

[26] T. Senthil *et al*, *Science* **303**, 1490 (2004).

[27] The 3D magnetic ordering wavevector has a finite component in the interchain direction due to antiferromagnetic couplings between chains.

[28] We acknowledge very useful discussions in particular with F.H.L. Essler and L. Balents, and also G. Mussardo, S.T. Carr, A.M. Tsvelik and M. Greiter. Work at Oxford, Bristol and ISIS was supported by the EPSRC UK and at HZB by the European Commission under the 6th Framework Programme through the Key Action: Strengthening the European Research Area, Research Infrastructures. Contract No: RII3-CT-2003-505925 (NMI3).

**Figures**

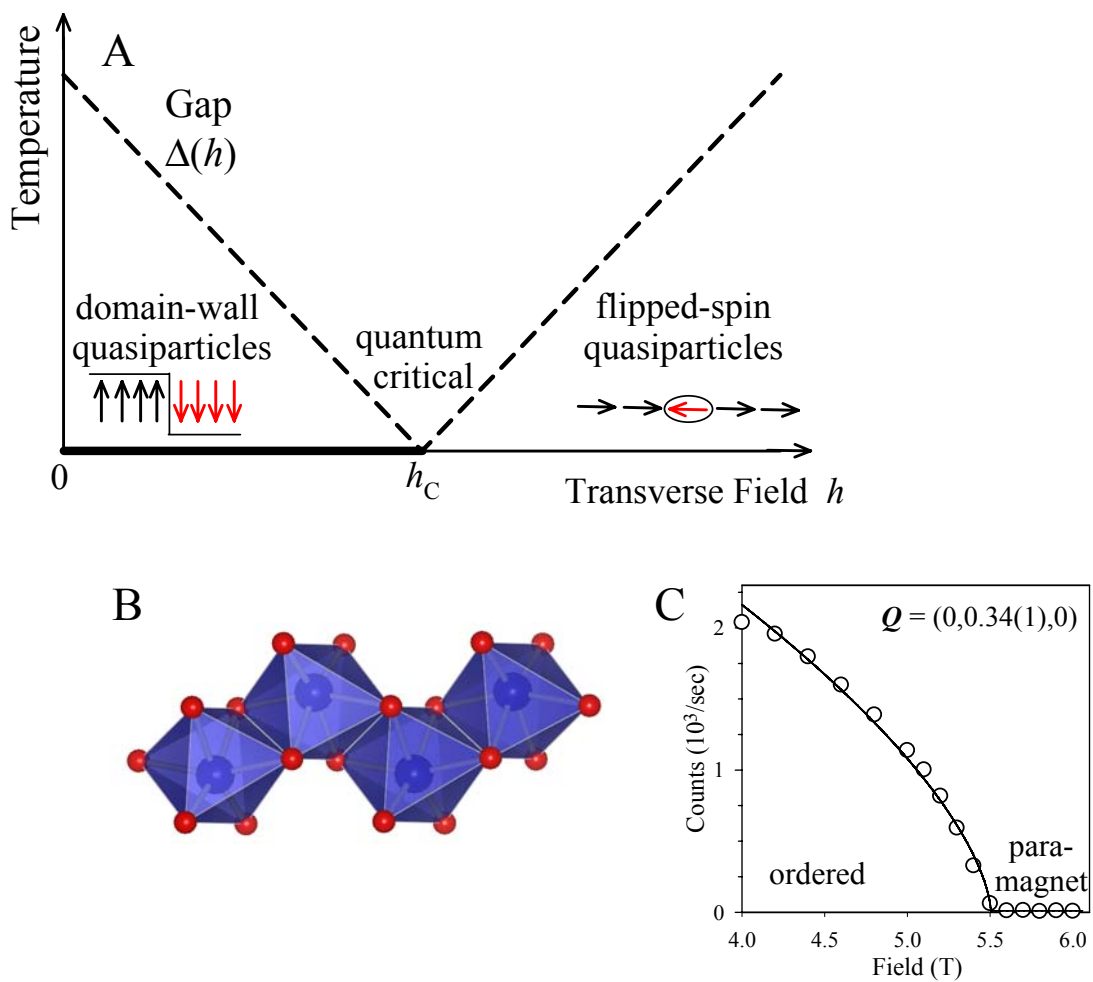

**Fig. 1** (A) Phase diagram of the Ising chain in transverse field Eq. 1. Spin excitations are pairs of domain-wall quasiparticles (kinks) in the ordered phase below $h_C$ and spin-flip quasiparticles in the paramagnetic phase above $h_C$. The dashed line shows the spin gap. (B) $CoNb_2O_6$ contains zig-zag ferromagnetic Ising chains. (C) Intensity of the 3D magnetic Bragg peak as a function of applied field observed by neutron diffraction (*27*).

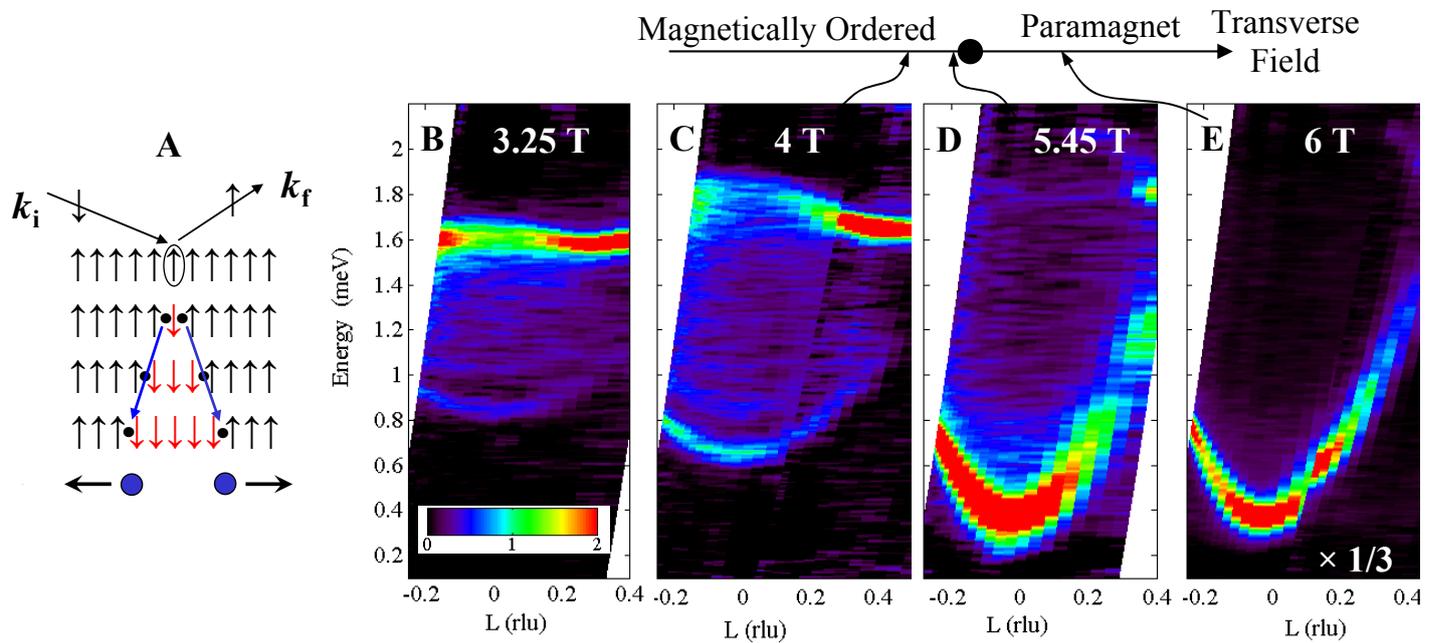

**Fig. 2** (A) Cartoon of a neutron spin-flip scattering that creates a pair of independently-propagating kinks in a ferromagnetically-ordered chain. (B)-(E) Spin excitations in $CoNb_2O_6$ near the critical field as a function of wavevector along the chain (in units of $2\pi/c$) and energy (*18*). In the ordered phase (B and C) excitations form a continuum due to scattering by pairs of kinks (as illustrated in A), in the paramagnetic phase (E) a single dominant sharp mode occurs, due to scattering by a spin-flip quasiparticle. Intensities in (E) are multiplied by 1/3 to make them comparable to the other panels.

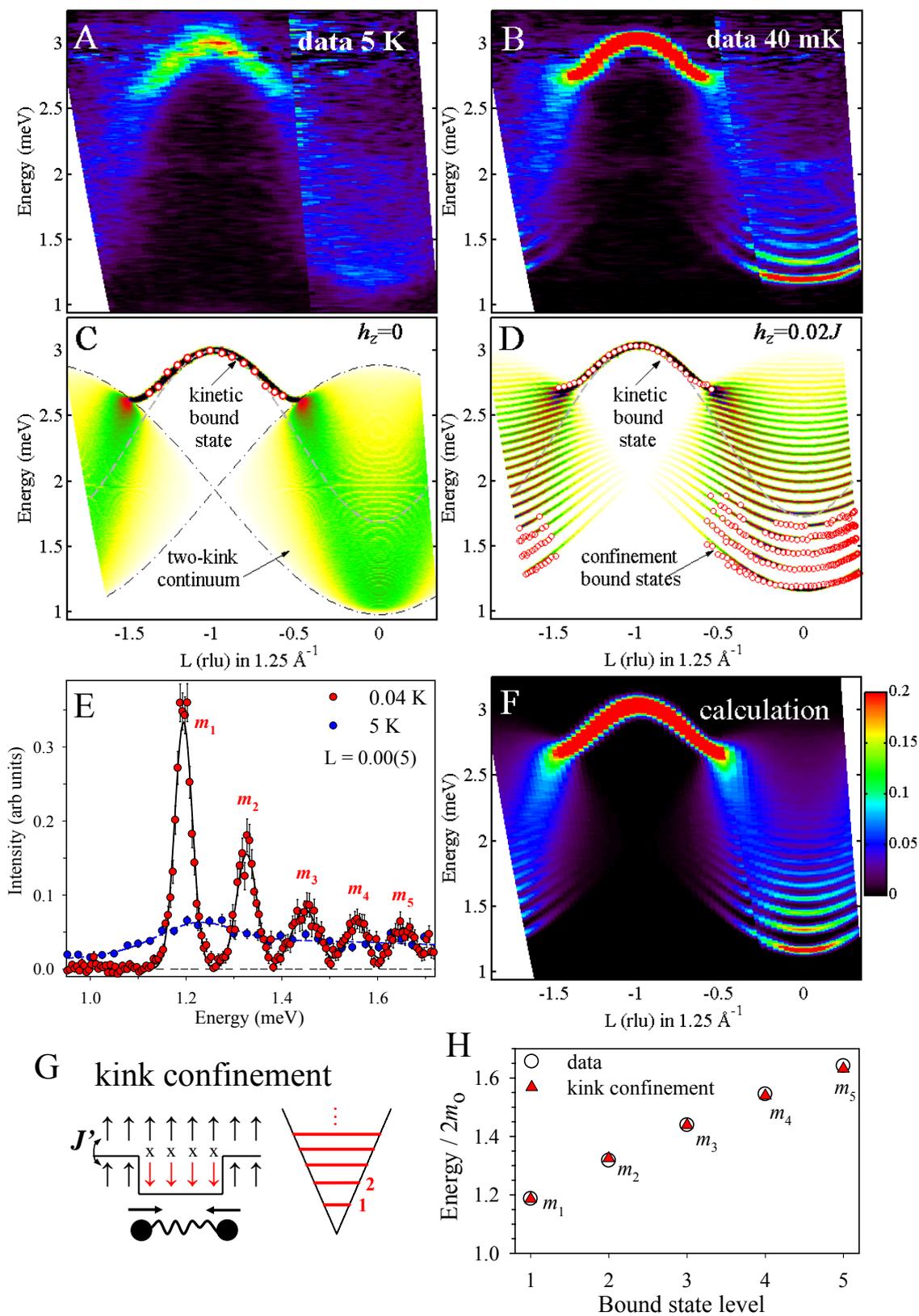

Figure 3

**Fig. 3** Zero-field spin excitations in $CoNb_2O_6$. (A) Above $T_N$ a broad continuum occurs near the zone centre (L=0) due to scattering by pairs of unbound kinks. (B) The continuum splits into a Zeeman ladder of two-kink bound states deep in the ordered phase. (C) and (D) Model calculations (Eq. S1) for $h_z$=0 and 0.02$J$ to compare with data in (A) and (B), respectively. In (C) the thick dashed line is the kinetic two-kink bound state stable only outside the two-kink continuum (bounded by the dashed-dotted lines). Open symbols in (C)/(D) are peak positions from (A)/(B). (E) Energy scan at the zone centre observing five sharp modes (red/blue circles are data from (B)/(A), solid line is a fit to Gaussians). (F) Dynamical correlations $S^{xx}(k,\omega)$ (Eq. S1) convolved with the instrumental resolution to compare with data in (B). (G) In the ordered phase kink separation costs energy as it breaks interchain bonds $J'$, leading to an effective linear "string tension" that confines kinks into bound states. (H) Observed and calculated bound state energies.

**Fig. 4** (A) and (B) Energy scans at the zone centre at 4.5 and 5 T observing two peaks, $m_1$ and $m_2$, at low energies. (C) Softening of the two energy gaps near the critical field (above ~5T the $m_2$ peak could no longer be resolved). Points come from data as in Fig. 2B-D, lines are guides to the eye. The incomplete gap softening is attributed to the interchain couplings as described in the text. (D) The ratio $m_2/m_1$ approaches the E8 golden ratio (dashed line) just below the critical field. (F) Gapless continuum of critical kinks (shaded area) predicted for the critical Ising chain. (G) E8 spectrum expected for finite $h_z$. Lines indicate bound states and shaded area is the $2m_1$ continuum. (E) Expected lineshape in the dominant dynamical correlations at the zone centre $S^{zz}(k = 0,\omega)$ for case (G) (vertical bars are quasiparticle weights (7) relative to $m_1$): two prominent modes followed by the $2m_1$ continuum (schematic dashed line), in strong resemblance to observed data in (A,B).

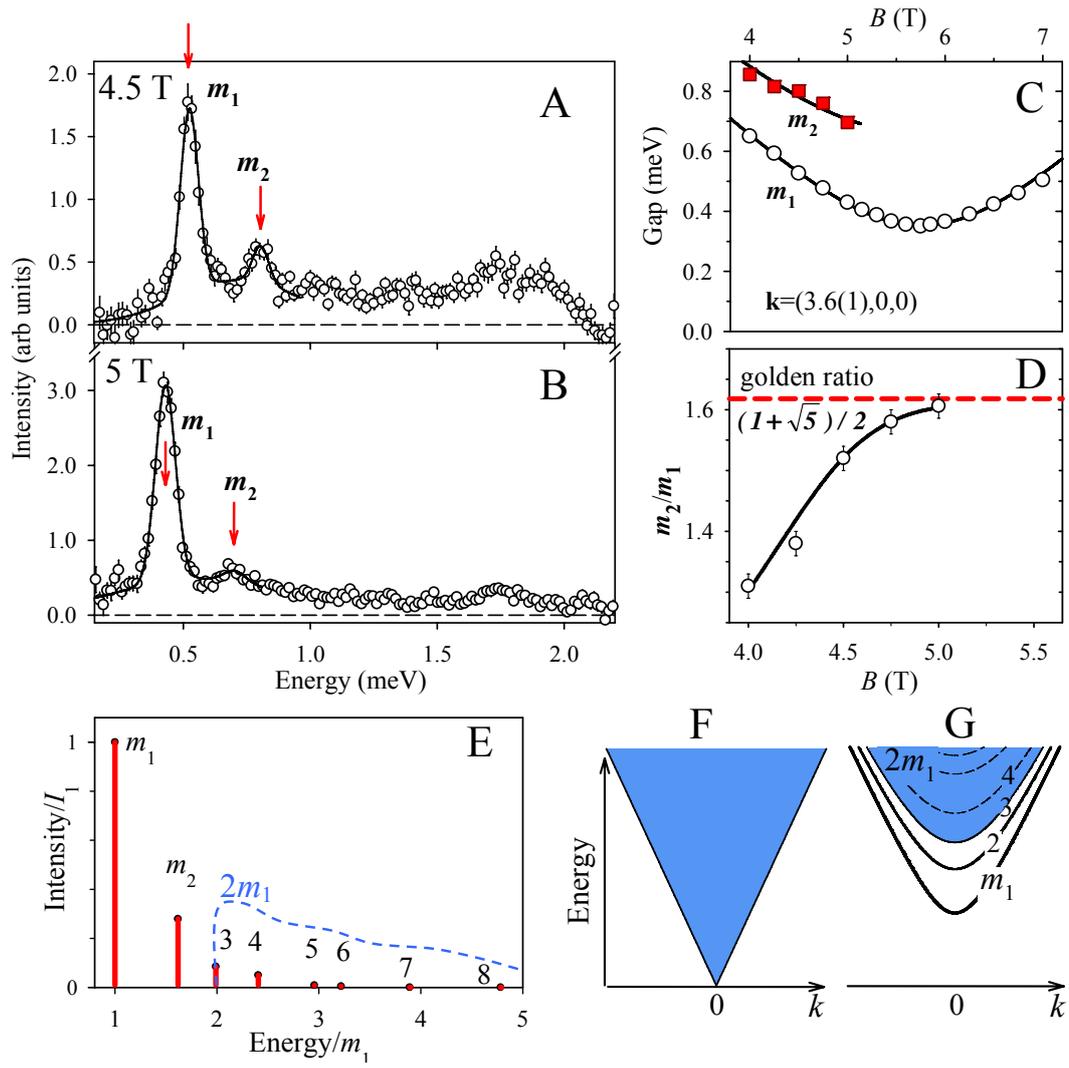

**Figure 4**

**Materials and methods**

**1. Neutron Scattering**

The zero-field inelastic neutron scattering data show in Fig. 3A-B was collected using the indirect-geometry time-of-flight spectrometer IRIS at the ISIS spallation neutron source at the Rutherford Appleton Laboratory in the UK. This allowed simultaneous measurements over a wide wavevector and energy range with high resolution, 0.021(1) meV (FWHM) on the elastic line using cooled pyrolytic graphite (002) analysers to select the final energy. The sample was an 8 gram, image furnace grown single crystal of $CoNb_2O_6$, cooled via a dilution refrigerator and mounted in the $[h,0,l]$ horizontal scattering plane. Data was collected at 0.04 K, 2.5 K, 5 K (1D phase) and 30 K (paramagnetic). For each temperature two sample orientations were measured with the chain direction (*c*-axis) parallel and rotated by 65º with respect to the incident beam direction $k_i$, this was chosen such that the covered wavevectors projected along the chain direction spanned a full Brillouin zone. Fig. 3A-B show the combined data from those two sample orientations, each counted for about 11 hours (~1800 µAmphrs). In Fig. 3A-B the color represents raw neutron counts with an estimate of the non-magnetic background subtracted off and divided by the magnetic form factor squared $f(Q)^2$ for $Co^{2+}$ ions and neutron polarization factor for fluctuations transverse to the Ising direction, as explained later in Sec 3.

Data in a magnetic field in Fig. 2B to E was collected using the related time-of-flight spectrometer OSIRIS, also at ISIS, with a similar size single crystal mounted in the same $[h,0,l]$ horizontal scattering plane. The sample was placed in a dilution refrigerator inside a vertical 7.5 T cryomagnet, such that magnetic fields were applied along the *b*-axis transverse to the Ising axes. The energy resolution on the elastic line was 0.027(1) meV (FWHM). Data was collected for a selection of magnetic fields up to 7 T, for each field typically a couple of sample orientations were measured (*c*-axis at 25º and 55º with respect to $k_i$), with each setting counted for about 5 hours (~800 µAmphrs). The colour plots in Figs. 2B-E are the raw time-of-flight neutron counts at T=0.1 K with an estimate of the non-magnetic background subtracted off.

The order parameter measurements in Fig. 1C were collected using the V2 cold-neutron triple-axis spectrometer at the Helmholz Centre Berlin, operated with neutrons of fixed final wavevector $k_F$=1.12 Å$^{-1}$, collimation guide-60'-open-open, vertically-focused PG002 monochromator and horizontally-focused PG002 analysers, this gave an energy resolution of 0.053(1) meV (FWHM) on the elastic line. The diffraction data in Fig. 1C was collected in elastic scattering mode with a 4.6 g single crystal mounted in the [0,k,l] horizontal scattering plane, placed inside a horizontal 6T cryomagnet aligned such as to have the magnetic field direction along the b-axis.

In both experiments in applied magnetic field the sample was held mechanically rigidly in place by a copper cage shaped around the crystal to avoid any possible movement of the sample in the presence of the large torques generated at high transverse fields.

## 2. Crystal structure and magnetism of $CoNb_2O_6$

$CoNb_2O_6$ crystallizes in the orthorhombic space group *Pbcn* with lattice parameters *a*=14.1337(1) Å, *b*=5.7019(2) Å, *c*= 5.0382(1) Å. The dominant magnetic interaction is ferromagnetic and occurs between nearest-neighbour $Co^{2+}$ spins arranged in zig-zag chains along the *c*-axis (Fig. 1B). Weak 3D couplings between chains stabilize long-range magnetic order below $T_{N1}$=2.95 K. Chains order ferromagnetically along their length with magnetic moments pointing along a local easy-axis (Ising) direction [contained in the crystal *ac* plane at $\gamma = \pm 29.6°$ to the *c*-axis, with $\pm$ for the two symmetry-equivalent chains in the unit cell (*14*)]. Two distinct 3D order patterns between chains occur as a function of temperature: for T<$T_{N2}$=1.97 K commensurate antiferromagnetic along *b* with $Q_{AF}$=(0,½,0), and for $T_{N2}$ <T<$T_{N1}$=2.95 K incommensurate spin-density wave along *b* with a temperature-dependent $Q$; this can be understood in terms of coupling of Ising chains by antiferromagnetic interchain couplings in a distorted triangular lattice in the *ab* plane (*14*). Throughout this paper wavevectors are expressed in reduced reciprocal lattice units of ($2\pi/a, 2\pi/b, 2\pi/c$).

## 3. Spin dynamics in zero-field: effective Hamiltonian for kink quasiparticles

The observed spectrum in Fig. 3B with co-existing continua and sharp modes motivates a phenomenological model of kinks with potential and kinetic energy terms. Near zero momentum such a problem can be modelled in terms of a Schrödinger's equation for two kinks as shown in Eq. 2, here we construct a generalization of this to finite wavevectors. We start from the Ising limit and add various perturbations that can lift the degeneracy of two-kink states and treat all effects to $1^{st}$ order in the perturbations. We restrict our attention to two-kink states only, i.e. $n$-spin cluster states $|i,n\rangle = |\uparrow\uparrow\downarrow_i\downarrow\ldots\downarrow\downarrow_{i+n-1}\uparrow\uparrow\rangle$ where the two indices give the start position and length of cluster, respectively, and consider the following effective Hamiltonian

$$\mathcal{H} |i, n\rangle = J |i,n\rangle - \alpha [|i,n+1\rangle + |i,n-1\rangle + |i+1,n-1\rangle + |i-1,n+1\rangle]$$
$$+ h_z\, n\, |i,n\rangle - \beta\, \delta_{n,1} [|\uparrow\downarrow_{i-1}\uparrow\uparrow\uparrow\rangle + |\uparrow\uparrow\uparrow\downarrow_{i+1}\uparrow\rangle] + \beta'\, \delta_{n,1} [|\uparrow\downarrow_i\uparrow\rangle] \qquad (S1)$$

Here the first term $J$ is the (dominant) energy cost to create two kinks in the absence of other perturbations, the $\alpha$ terms are an effective parameterization of the hopping of kinks left and right along the chain required to explain the observed band-width of the continuum near the zone centre (L=0) [there is an implicit repulsion between kinks due to the Pauli principle preventing double occupation of the same site, i.e. the terms $|i,n-1\rangle$ and $|i+1,n-1\rangle$ are implicitly absent for $n=1$]. $h_z$ is the effective longitudinal mean-field due to interchain couplings that occurs in the ordered phase when $\langle S^z \rangle \neq 0$, and leads to kink confinement. The $\beta$ term is a short-range interaction manifested in a kinetic energy gain for hopping for nearest-neighbour kinks; this stabilizes a spin-wave like bound state near the zone boundary (L=-1) ($\beta$ tunes the curvature of the dispersion and $\beta'$ shifts the absolute energy $\omega_{L=-1}=J+h_z+2\beta+\beta'$).

The spin excitation spectrum of the Hamiltonian in Eq. S1 in momentum space can be obtained by diagonalization in the basis of Fourier-transformed spin-cluster states (*S1*). The connection with Schrödinger's equation applicable near zero momentum becomes apparent when considering the case

$\beta=\beta'=h_z=0$ when the model has free kinks with a quadratic dispersion near the band minimum $\varepsilon(k) = m_o + \hbar^2 k^2/(2\mu)$ with $m_o = J(1-4\alpha)/2$, $\hbar^2/(2\mu\tilde{c}^2) = \alpha J$ and momentum $k=\pi L/\tilde{c}$. A finite field $h_z$ is equivalent to a string tension between two kinks so the problem maps onto Schrödinger's Eq. 2 with $\lambda = h_z/\tilde{c}$. The experimentally-observed sharp peak energies are reproduced very well by the expected bound state spectrum (Fig 3H), with the fit parameters to Eq. 3 obtained as $m_o$ = 0.503(5) meV and $\lambda\tilde{c}$ =0.033(1) meV. Here the one-kink kinetic energy prefactor $\hbar^2/(2\mu\tilde{c}^2)$ = 0.23(2) meV was obtained separately from a global fit to the dispersions at finite wavevector. For finite momentum $k$ away from the zone centre the bound state spectrum is obtained as a generalization of Eq. 3 as $m_j(k) = 2\varepsilon(k/2) + z_j \lambda^{2/3} (\hbar^2/\mu)^{1/3}$, $j$=1,2,3…, i.e. the kinetic energy of the composite two-kink bound state is equal to the kinetic energy of the two constituent kinks each carrying half of the total momentum (*22*).

The above Eq. S1 is motivated empirically whereby each term has a specific role to explain features in the data. Some terms are known to arise in the effective description of the low-energy dynamics of other Ising-like magnets, for example CsCoCl$_3$ (*S1*), where it is well understood how projection of the Heisenberg magnetic exchange between neighbouring Co$^{2+}$ ions onto the doublet ground state of the (hexagonal symmetry) crystal field and spin-orbit coupling Hamiltonian of a Co$^{2+}$ ion leads to an effective spin-1/2 Hamiltonian with a dominant Ising exchange $JS_i^z S_j^z$ and additional smaller transverse couplings of the form $2\beta(S_i^x S_{i+1}^x + S_i^y S_{i+1}^y) = \beta(S_i^+ S_{i+1}^- + S_i^- S_{i+1}^+)$. We note that the microscopic origin of the additional terms in Eq. S1 apart from $J$ and $\beta$ may be related to the very low local symmetry of the Co$^{2+}$ ions (monoclinic $C_2$ point group), which may allow additional coupling terms that cancel in higher-symmetry crystal-field environments, also dipolar effects or further-neighbour couplings along the chain could be important (*S2*).

The neutron-scattering intensity is proportional to the wavevector and energy-dependent transverse dynamical correlation functions $S^{xx}(k,\omega) = S^{yy}(k,\omega) = \Sigma_{|E\rangle} |\langle E|S^x(k)|0\rangle|^2 \delta(\omega-\omega_E)$, where the sum extends over all excited eigenstates |E⟩ of momentum $k$ and energy $\omega_E$ relative to the ground state |0⟩. The calculated spin excitation spectrum obtained this way is shown in Fig. 3F, to be directly compared with the data in Fig. 3B (*S3*). The parameters are $J = 1.94(4)$ meV, $\alpha = 0.12(1)\,J$, $h_z = 0.020(2)\,J$, $\beta = 0.17(1)\,J$, $\beta' = 0.21(1)\,J$. These values were obtained by a best global fit to the experimentally extracted dispersions of the five bound states near L=0, the sharp mode dispersion near the zone boundary L=-1, and the relative intensity ratio of those modes, with equal weight in the fits given to all those features.

**References and Notes**

[S1] J. P. Goff, D. A. Tennant, S. E. Nagler, *Phys. Rev. B* **52**, 15992 (1995).

[S2] The presence of the finite term $\beta'$ which indicates a different potential energy for the two domain walls when they are one site apart (energy $J+\beta'$) compared to when they are two or more sites apart (energy $J$), could be explained by the presence of a finite second-neighbor Ising antiferromagnetic coupling along the chains $J_2/J_1 = -\beta'/(J+2\beta') = -15\%$, which may be due to superexchange via the Co-O-O-Co path, possible given the zig-zag chain geometry shown in Fig. 1B. Here the nearest-neighbor Ising coupling is $J_1 = J+2\beta'$.

[S3] The intensities in Fig. 3A-B have been corrected for the neutron polarization factor as follows. Near the Ising limit the dominant spin fluctuations are polarized transverse to the Ising axis with the intensity in neutron scattering proportional to $p(\mathbf{u})\,S^{xx}(\mathbf{k},\omega)$ where we have assumed: i) $S^{xx}(\mathbf{k},\omega)=S^{yy}(\mathbf{k},\omega)$, i.e. isotropic fluctuations in both directions transverse to the Ising axis and ii) negligible weight at finite energy in the other dynamical correlation $S^{zz}$. The polarization factor is $p(\mathbf{u})=1+(\hat{\mathbf{k}}\cdot\mathbf{u})^2$ where $\hat{\mathbf{k}}$ is the unit vector defining the direction of the neutron wavevector transfer $\mathbf{k}$. To take into account the two Ising chains in the unit cell with rotated easy axes $\mathbf{u}_{1,2}=(\pm\sin\gamma,0,\cos\gamma)$, the neutron data was divided by the average $[p(\mathbf{u}_1)+p(\mathbf{u}_2)]/2$.